\newcommand{\CLASSINPUTinnersidemargin}{16.9mm}
\newcommand{\CLASSINPUTtoptextmargin}{19mm}
\newcommand{\CLASSINPUTbottomtextmargin}{25mm}
\documentclass[conference,a4paper]{IEEEtran}
\ifCLASSINFOpdf
   \usepackage[pdftex]{graphicx}
   \graphicspath{{./}}
   \DeclareGraphicsExtensions{.pdf,.jpeg,.png,.jpg}
\else
\fi
\usepackage[cmex10]{amsmath}
\usepackage{booktabs} 

\usepackage[caption=false,font=footnotesize]{subfig}
\usepackage[
pdfa,
hidelinks,
breaklinks=true,
bookmarksnumbered=true,
pdfpagemode=UseOutlines,
pdfpagelayout=SinglePage,
pdffitwindow=true,
bookmarksopen,
]{hyperref}
\RequirePackage[
nameinlink,
]{cleveref}
\newcommand{\email}[1]{\ttfamily\href{mailto:#1}{#1}}

\newlength{\MaxSizeOfLineNumbers}%
\usepackage{jlcode}
\newlength\columnlistingwidth
\usepackage[subdued,defaultmathsizes]{mathastext}
\renewcommand{\texttt}[1]{\ensuremath{\mathtt{#1}}}
\addlitjlbase{SOmega}{$\mathtt{\Omega}$}{1}
\addlitjlbase{Somega}{$\mathtt{\omega}$}{1}
\addlitjlbase{Sphi}{$\mathtt{\varphi}$}{1}

\newcommand{\pd}{PowerDynamics.jl}
\newcommand{\diffeq}{DifferentialEquations.jl}
\newcommand{\sundials}{SUNDIALS}
\newcommand{\LY}[1]{\ensuremath{Y^L_{#1}}}

\newcommand{\dynamicnodemacro}{\texttt{@DynamicNode}}

\newcommand{\nth}[1]{\ensuremath{#1^{\text{th}}}}
\renewcommand{\Re}{\ensuremath{\text{Re}}}
\renewcommand{\Im}{\ensuremath{\text{Im}}}

\begin{document}

\title{Sneak Preview: PowerDynamics.jl\\ -- \\An Open-Source library for analyzing dynamic stability in power grids with high shares of renewable energy}

\author{\IEEEauthorblockN{Tim Kittel, Sabine Auer and Christina Horn}
\IEEEauthorblockA{Potsdam Institute for Climate Impact Research (PIK)\\
Member of the Leibniz Association\\
P.O. Box 60 12 03\\
D-14412 Potsdam\\
Germany\\
Email: \email{tim.kittel@pik-potsdam.de}}}


%


\maketitle

\thispagestyle{plain}
\pagestyle{plain}

\begin{abstract}
PowerDynamics.jl is an Open-Source library for dynamic power grid modeling built in the latest scientific programming language, Julia. It provides all the tools necessary to analyze the dynamical stability of power grids with high share of renewable energy. In contrast to conventional tools, it makes full use of the simplicity and generality that Julia combines with highly-optimized just-in-time compiled Code. Additionally, its ecosystem provides DifferentialEquations.jl, a high-performance library for solving differential equations with built-in solvers and interfaces to industrial grade solvers like Sundials. PowerDynamics.jl provides a multitude of dynamics for different node/bus-types, e.g. rotating masses, droop-control in inverters, and is able to explicitly model time delays of inverters. Furthermore, it includes realistic models of fluctuations from renewable energy sources. In this paper, we demonstrate how to use PowerDynamics.jl for the IEEE 14-bus distribution grid feeder.
\end{abstract}


%
\IEEEpeerreviewmaketitle

\section{Introduction}


Dynamic stability analysis of power grids concerns the investigation of transient stability and power quality. Especially with more intermittent renewable generation sudden changes in power followed by the transient reaction of frequency and voltage dynamics may present a challenge to power system stability. At the same time, fluctuations also challenge power quality, which refers to whether voltage and frequency stay within safety bounds and whether the waveform has undesired harmonic distortions. 
Furthermore, the wide deployment of power electronics, witch RES are connected to the grid, introduces new features such as delayed reaction into future power grids.
 
These challenges have been the topic of recent research on the stability of renewable power grids \cite{auer2017stability,auer2018stability,schafer2016taming}. However, the usage of state-of-the-art Open-Source software repeatedly posed limitations to the possible modeling scope. One reason is the lack of solvers for the different types of dynamics which need to be described a stochastic differential equations, delay differential equations and differential algebraic equations. At the same time to network size was constrained by computational performance. This means a realistic model implementation and hence reliable stability analysis has been constrained by the available OS software framework. 

In this conference paper, we give a sneak preview to \pd{}, an new Open-Source approach to dynamic power grid modeling\footnote{This paper is meant as a sneak preview, as the planned publication date is Oct 15, 2018, i.e.\ before the Wind Integration Workshop 2018 will take place.}.
While there are multiple modeling suites available they are either proprietary (e.g.\  PSS\textregistered{} \cite{PSS}) or need proprietary tools to be used (e.g.\ PSAT \cite{PSAT, milano2008open}). With \pd{}, we aim to provide a modeling framework that tackles all the needs a modeler for power grids with high shares of renewables to execute dynamic analyses and the first publication will provide tools for time-domain modeling.

At the moment, \pd{} is under high development, with the basic software architecture already being developed and implemented. We are currently adding new types of dynamics and controls to represent different buses, e.g. droop control, multiple inverter types, higher-order representations of synchronous machines, and in particular intermittency representations of renewable energy sources. A second focus of effort goes into usability, in order to improve the experience of modelers and decrease their time they have to concern with software-related problems so they can focus on their modeling instead. Our vision is to provide a modular, fully flexible library where modelers can decide whether they (a) want to simply use repdefined bus models to analyze a power grid quickly, (b) want to into all the detail of modeling each bus type, or (c) want to find their right position in-between (a) and (b). All of that is done in the context of an Open-Source context in order to make it available to and invitre contributions from research and industry.

\pd{} is written in Julia \cite{bezanson2017julia}, a new, fast developing programming language targeting the scientific computing community. With its recent release to Julia 1.0, it has matured enough to become a powerful player competing with more traditional programming languages like Python and C in the modeling world.
In the wider context, there is a current movement within the Open Energy Modeling community \cite{openmod} to switch to Julia and unite modeling efforts. With \pd{} we are aiming to support these efforts by contributing to dynamic power system analysis.

The rest of the paper is structured as follows: in \Cref{sec:julia} ``\nameref{sec:julia}'' we give a detailed reasoning why we find Julia to be the correct choice as the programming language for our modeling effort; in \Cref{sec:powerdynamics} ``\nameref{sec:powerdynamics}'' we introduce the basic mathematical notions we use in \pd{}; in \Cref{sec:ieee14} ``\nameref{sec:ieee14}'' we provide the details of the example system used in this paper; in \Cref{sec:implementation} ``\nameref{sec:implementation}'' we show how to implement the IEEE 14-bus feeder in \pd{}; in \Cref{sec:modeling} ``\nameref{sec:modeling}'' we present the results for our example model; and in \Cref{sec:conclusion} ``\nameref{sec:conclusion}'' we conclude the paper.

\subsection*{Convention \& Notation}

Within this paper, we take up the convention of \pd{} where minor letters are used as variables for dynamic variables and capital letters denote (constant) parameters and all variables are in defined in the co-rotating frame.

Further, we take $u_a$ to to be the complex voltage, $v_a = |u_a|$ to be the voltage magnitude, $\varphi_a = \arg(u_a)$ to be the voltage angle, $i_a = \sum_b \LY{ab}u_b$ to be the complex current, $s_a = u_a \cdot i_a^*$ to be the complex power, $p_a = \Re(s_a)$ the active power and $q_a = \Im(s_a)$ to be the reactive power of the \nth{$a$} bus. \LY{} is the admittance Laplacian. The imaginary element is denoted as $j = \sqrt{-1}$.

\section{Julia for Scientific Modeling}
\label{sec:julia}

The advantages of Julia for scientific modeling can be summarized as (adapted from \cite{julia}):
\begin{itemize}
	\item \textbf{Julia is fast!} It was designed from the beginning as a high-performance language. Using the just-in-time (JIT) compilation method, it produces efficient native code for multiple platforms.
	\item \textbf{Julia allows for rapid development!} It uses dynamic typing, so it is easy to use, feels like a scripting language and has good interactive use without losing any performance. That way, it saves the developer a lot of time.
	\item \textbf{Julia is technical!} It is developed for numerical computing, the syntax is focusing on precise mathematics and many datatypes and even parallelism is available out of the box.
	\item \textbf{Julia is general!} Using multiple dispatch as the fundamental paradigm, it is easy to express object-oriented and functional programming patterns at the same time.
	\item \textbf{Julia is composable!} Julia has been designed such that independent packages work well together. Hence, we can use matrices of unit quantities or differentiation with sparse matrices without having these types ever defined explicitly before.
	\item \textbf{Julia has a growing ecosystem!} The aforementioned advantages attract a growing community\footnote{The community growth recently reached the point where the organizers had to cut off ticket sales for this year's JuliaCon at some point.} and with that a growing ecosystem of Julia packages. Particularly important for modeling is the \diffeq{} library which provides high-performance solvers and interfaces to industrial grade solvers like \sundials{} \cite{hindmarsh2005sundials} for solving multiple kinds of differential equations (ordinary, algebraic, delayed, stochastic).
	\item \textbf{Julia allows for metaprogramming!} This means, at execution time, the source code is available as data and one can easily modify it. Furthermore, one can even add own syntax to Julia. Within \pd{}, we make heavy use of metaprogramming to make it as simple as possible for a user to implement her/his own kind of bus model (see \Cref{sec:implementation-nodes}). Also, it allows to write very simple code that can automatically be modified to highly optimized code.
\end{itemize}

Of course, this is just a simply overview and we would strongly recommend the reader to try out Julia and convince herself/himself\footnote{\url{https://julialang.org/learning/}}.

\section{Mathematics of \pd{}}
\label{sec:powerdynamics}

Within this paper, we take the approach to describe the dynamics of a power grid by a (set of) semi-explicit differential algebraic equation(s). We explicitly distinguish between the complex voltages $u_a$ (of the \nth{$a$} bus), whose information is transmitted between buses directly through the complex current $i$, and internal variables $x_{ab}$ (the \nth{$b$} internal variable of the \nth{$a$} bus), that exist locally at a node. An example for an internal variable would be the frequency of a synchronous machine. While the phase angle $\varphi$ is transmitted directly via the current, its derivative is the angle velocity whose dynamics is defined locally by the the rotating mass. We are aware that a multitude of approaches are possible to write down the dynamics and we have chosen this particular one as it can be easily translated into source code. Writing this idea in formulas yields
\begin{subequations}
\begin{align}
i_a = \sum_c \LY{ac} \cdot u_c , \label{eq:i-definition} \\
m^u_a \dot u_a = f_a(u_a, x_a, i_a)  ,\label{eq:u-general} \\
m^x_{ab} \dot x_{ab} = g_{ab}(u_a, x_a, i_a) , \label{eq:x-general}
\end{align}
\end{subequations}
where $m^u$ / $m^x$ are the masses for the voltages / internal variables respectively\footnote{Within \pd{} we allow masses to be $1$ or $0$.}. Also, $f_a$ encodes the voltage dynamics of the \nth{$a$} bus and $g_{ab}$ the dynamics of the \nth{$b$} internal variable of the \nth{$a$} bus. Note the convention of using minor letters for dynamical variables.

\LY{} is the admittance Laplacian. So having the admittances encoded as $Y_{ab}$ where $Y_{ab} = 0$ if there is no line, or $Y_{ab}$ with a complex value (with $\Re(Y_{ab})\geq 0$) for existing lines, then the admittance Laplacian is defined as 
\begin{align}
\LY{ab} = \delta_{ab}\sum_c Y_{ac} - Y_{ab}, 
\end{align}
where $\delta_{ab}$ is Kronecker-$\delta$.
Particular examples would be an algebraic slack bus as the \nth{$a$} bus with a fixed complex voltage $U_a$. There are no internal variables and the voltage mass is $m_a^u = 0$ (as its an algebraic equation), so it can simply be written as
\begin{align}
0\cdot \dot u_a = f_a(u_a, x_a, i_a) = u_a - U_a , \label{eq:slack-bus}
\end{align}
where we have explicitly not removed $\dot u$ in order to keep a strong analogy the the implementation shown later.
Similarly, we can treat an algebraic load at the \nth{$a$} bus with a constant complex power $-S_a$ flowing out\footnote{The $-$ is due to the definition of all variables such that the power flows always from the respective node into the power grid.}. Again, there are no internal variables and the voltage mass is $0$ so it reduces to
\begin{align}
0\cdot \dot u_a = f_a(u_a, x_a, i_a) = s_a - S_a = u_a\cdot i_a^* - S_a .
\end{align}
A synchronous machine as the \nth{$a$} bus can be represented by the Swing equation (or $2^{\text{nd}}$-order synchronous machine model) \cite{sauer1998power,kundur1994power} with a produced active power $P_a$, the damping constant $D_a$, the rated frequency $\Omega$ and the inertia $H_a$. In this case, the voltage mass is $1$, there is one internal variable, the frequency, $x_{a1} = \omega_a$ with a mass of $1$ because it is dynamic and not an algebraic constraint\footnote{In vector notation this gives $x_{a} = (\omega_a)$ for the Swing equation.}. Hence, the equations reduce to
\begin{subequations}
\begin{align}
&\dot u_a = f_a(u_a, (\omega_a), i_a) = u_aj\omega_a , \label{eq:swing-u}  \\
&\begin{aligned}
\dot x_{a1} = \dot\omega_a &= g_{a1}(u_a, (\omega_a), i_a)\\
&= \frac{2\pi\Omega}{H_a}\left( P_a - D_a\omega_a - \Re(u_a\cdot i_a^*)  \right) 
\end{aligned} \label{eq:swing-omega}
\end{align}
\end{subequations}
where we would like to remind the reader of $j$ being used as the imaginary element in this paper. Note that with $u_a = e^{j\varphi_a}\ \implies\ du_a = ju_ad\varphi_a$, \Cref{eq:swing-u} reduces to $\dot\varphi_a = \omega_a$ and one recovers the usual version in terms of the voltage angle.

In the following section, we will introduce the example system in order to demonstrate afterwards, how these kind of dynamics are then described in \pd{}.

\section{The IEEE 14-bus distribution grid feeder}
\label{sec:ieee14}

\begin{figure}[!t]
	\centering
	\includegraphics[width=\columnwidth]{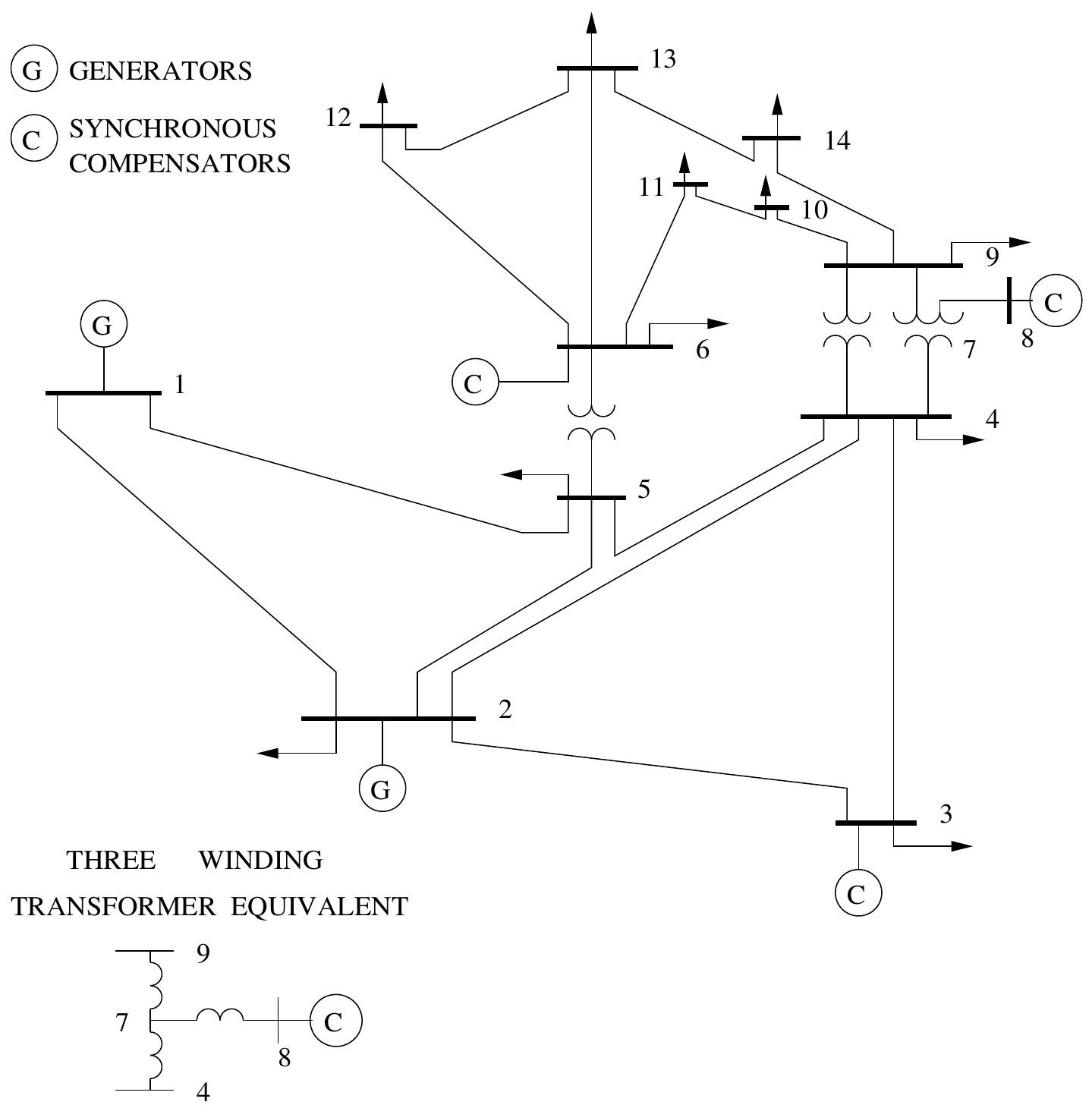}
	\caption{Technical representation of the IEEE 14-bus test system, extracted from \cite{kodsi2003modeling}. It contains five synchronous machines of which two are generators. The one at bus 2 is taken as the slack bus for the modeling in this paper. Furthermore, there are 8 loads (buses 4, 5, 8, 9, 10, 11, 12, 13, and 14). Bus 7 is for the representation of the transformer.}
	\label{fig:ieee14grid}
\end{figure}

The IEEE 14-bus system is a representation of a medium-voltage distribution grid. A graphical representation is in \Cref{fig:ieee14grid}, which is extracted from \cite{kodsi2003modeling}. There are two generators, where the one at bus 2 is the slack bus, three synchronous compensators, and eight (complex) loads. Bus 7 is for the representation of the transformer only.

The generator at bus 1 and the synchronous compensators will be modeled with the swing equation, the slack bus 2 with the algebraic constraint introduced in \Cref{sec:powerdynamics} and the loads with the algebraic constraints for complex loads (see \Cref{sec:powerdynamics} also). The bus parameters haven been extracted from \cite{kodsi2003modeling} and are presented in \Cref{tab:ieee14-bus-parameters}. The resistance $R$ and reactance $X$ for the lines (taken from \cite{kodsi2003modeling} as well) are summarized in \Cref{tab:ieee14-line-parameters}.

Having a description of the power grid in question enables us to have a look at the implementation in \pd{} now.

\setlength\tabcolsep{1.5mm}
\renewcommand{\arraystretch}{1.2}
\begin{table}[!t]
\caption{Bus parameters of the IEEE 14-bus test system. INERTIA UNIT IS?}
\label{tab:ieee14-bus-parameters}
\centering
\begin{tabular}{|c|c||c|c|c|c|c|}
\toprule
Bus & Type & $U\,[p.u.]$ & $P\,[p.u.]$ & $Q\,[p.u.]$ & $D\,[p.u.]$ & $H[p.u.\cdot s]$ \\\midrule
1 & Generator &  & 2.32 & & 2 & 5.148\\\hline
2 & Slack & 1 & & & & \\\hline
3 & Syn. Comp. & & -0.942 & & 2 & 6.54 \\\hline
4 & Load & & -0.478 & 0 & & \\\hline
5 & Load & & -0.076 & -0.016 & & \\\hline
6 & Syn. Comp. & & -0.112 & & 2 & 5.06 \\\hline
7 & Load & & 0 & 0 & & \\\hline
8 & Syn. Comp. & & 0 & & 2 & 5.06 \\\hline
9 & Load & & -0.295 & -0.166 & & \\\hline
10 & Load & & -0.09 & -0.058 & & \\\hline
11 & Load & & -0.035 & -0.018 & & \\\hline
12 & Load & & -0.061 & -0.016 & & \\\hline
13 & Load & & -0.135 & -0.058 & & \\\hline
14 & Load & & -0.149 & -0.05 & & \\
\bottomrule
\end{tabular}
\end{table}

\begin{table}[!t]
\caption{Line parameters of the IEEE 14-bus test system.}
\label{tab:ieee14-line-parameters}
\centering
\begin{tabular}{|c||c|c|c|c|}
\toprule
line number & from bus & to bus & $ R\, [p.u.]$ & $ X\, [p.u.]$ \\\midrule1 & 1 & 2 & 0.01938 & 0.05917\\\hline2 & 1 & 5 & 0.05403 & 0.22304\\\hline3 & 2 & 3 & 0.04699 &0.19797\\\hline4 & 2 & 4 & 0.05811 & 0.17632\\\hline5 & 2 & 5 & 0.05695 & 0.17388\\\hline6 & 3 & 4 & 0.06701 & 0.17103\\\hline7 & 4 & 5 & 0.01335 & 0.04211\\\hline8 & 4 & 7 & 0.0 & 0.20912\\\hline9 & 4 & 9 & 0.0 & 0.55618\\\hline10 & 5 & 6 & 0.0 & 0.25202\\\hline11 & 6 & 11 & 0.09498 & 0.1989\\\hline12 & 6 & 12 & 0.12291 & 0.25581\\\hline13 & 6 & 13 & 0.06615 & 0.13027\\\hline14 & 7 & 8 & 0.0 & 0.17615\\\hline15 & 7 & 9 & 0.0 & 0.11001\\\hline16 & 9 & 10 & 0.03181 & 0.0845\\\hline17 & 9 & 14 & 0.12711 & 0.27038\\\hline18 & 10 & 11 & 0.08205 & 0.19207\\\hline19 & 12 & 13 & 0.22092 & 0.19988\\\hline20 & 13 & 14 & 0.17093 & 0.34802\\
\bottomrule
\end{tabular}
\end{table}

\section{Implementation in PowerDynamics.jl}
\label{sec:implementation}

In this section, we will show first how to implement the dynamics for the different buses presented in \Cref{sec:powerdynamics} in \pd{} and then how create a power grid model from these bus dynamics definitions, in this case the IEEE 14-bus system.

\subsection{Defining the dynamics}
\label{sec:implementation-nodes}

The implementation of such a power grid in \pd{} is strongly based on the \dynamicnodemacro{} macro. A macro is a metaprogramming function that takes part of the source codes and modifies it at parsing time before giving it to the compiler, that creates the actual bitcode. In case of \dynamicnodemacro{}, this is used to hide all the complicated internals of \pd{} and make the definition of a type of dynamics as easy as possible. The following examples are already provided by default in \pd{} but using them as an example, any kind of dynamics can be implemented.

Using \dynamicnodemacro{}, the slack bus constraint from \Cref{eq:slack-bus} is implemented as
\begin{lstlisting}[linewidth=\columnlistingwidth]
@DynamicNode SlackAlgebraic(U) <: OrdinaryNodeDynamicsWithMass(m_u=false) begin
end [] begin
    du = u - U
end
\end{lstlisting}
where line breaks and line numbers have been added for easier orientation.
The left part on the first line states the name of the new type \texttt{SlackAlgebraic} and the parameter name \texttt{U}. Separated by the subtyping operator (\texttt{<:}) comes the statement that it can be represented by an ordinary differential equation with a mass term, where the mass for the voltage \verb|m_u| is set to \texttt{false}, meaning $m^u = 0$ from \Cref{eq:u-general} as it is a constraint.  The \verb|[]| in line 2 states that there are no internal variables. And line 3 simply provides the formula for the dynamics and assigns it to \texttt{du}. As \verb|m_u = false| is given, this simply translates it to a constraint.

In a similar manner, the load is defined as an algebraic PQ-constraint
\begin{lstlisting}[linewidth=\columnlistingwidth]
@DynamicNode PQAlgebraic(S) <: OrdinaryNodeDynamicsWithMass(m_u=false)  begin
end [] begin
    s = u*conj(i)
    du = S - s
end
\end{lstlisting}
In this case, an additional line was added to calculate the outflowing complex power \texttt{s} from the voltage \texttt{u} and the current \texttt{i} and then the definition of the constraint is written in line 4.

Concerning the generator and the synchronous compensators, we implemented the swing equation as
\begin{lstlisting}[linewidth=\columnlistingwidth]
@DynamicNode SwingEq(H, P, D, SOmega) <: OrdinaryNodeDynamics() begin
	@assert D > 0 "damping (D) should be >0"
	@assert H > 0 "inertia (H) should be >0" 
	SOmega_H = SOmega * 2pi / H
end [[Somega, dSomega]] begin
	p = real(u * conj(i))
	du = u * im * Somega
	dSomega = (P - D*Somega - p)*SOmega_H
end
\end{lstlisting}
Again, the left part of the first line provides the name of the new type and the name of the parameters. Lines 2 to 4 are code that should be run only once. In this case these are consistency checks, that the damping and inertia are positive, and the reduction of the rated frequency $\mathtt{\Omega}$ and the inertia $\mathtt{H}$ to a single variable $\mathtt{\Omega\_H}$. In line 5, the variable name of the internal variable $\mathtt{\omega}$ and the name for its derivative $\mathtt{d\omega}$ are given. Finally, lines 6 to 8 implement \Cref{eq:swing-u,eq:swing-omega} by simply writing down the mathematical terms.

\subsection{Instantiating the power grid model}
\label{sec:implementation-grid}

In order to create the grid model, we first have to instantiate the bus models simply by calling them with the corresponding parameter values from \Cref{tab:ieee14-bus-parameters}, e.g.:
\begin{lstlisting}[linewidth=\columnlistingwidth]
SwingEq(H=5.148, P=2.32, D=2, SOmega=50) # for bus 1
PQAlgebraic(S=-0.295-0.166im) # for bus 9
\end{lstlisting}
Within the actual code we simply loaded the data from a \texttt{.csv} file into and automatized this instantiation\footnote{The full source code is not part of this paper as it would be too long, but it will be published along with \pd{}.}. The instantiated bus models should then be saved in an array called e.g. \texttt{nodes}. Similarly, the admittance Laplacian should be generated from the line data in \Cref{tab:ieee14-line-parameters} and saved in a matrix called e.g. \texttt{LY}. The actual grid model instatiation is then simply one line where the model is saved in the variable \texttt{g}:
\begin{lstlisting}[linewidth=\columnlistingwidth]
g = GridDynamics(node_list, LY)
\end{lstlisting}
Now, \texttt{g} contains all the information of the power grid and in the following section we will show how to solve it.

\section{Modeling Results}
\label{sec:modeling}

\begin{figure*}[!t]
	\centering
	\subfloat{\includegraphics[width=\columnwidth]{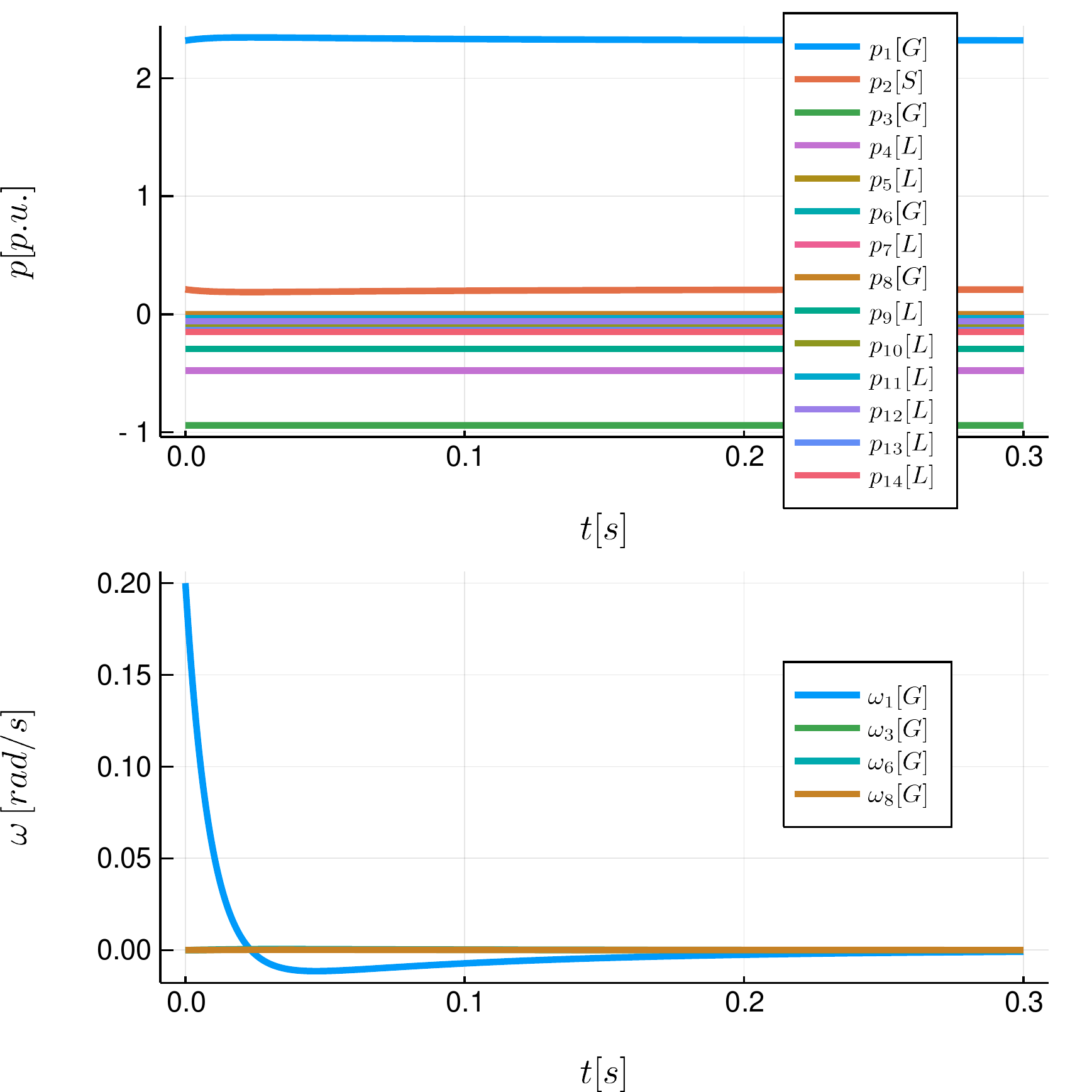}
	\label{fig:ieee14-frequency-perturbation}}
	\subfloat{\includegraphics[width=\columnwidth]{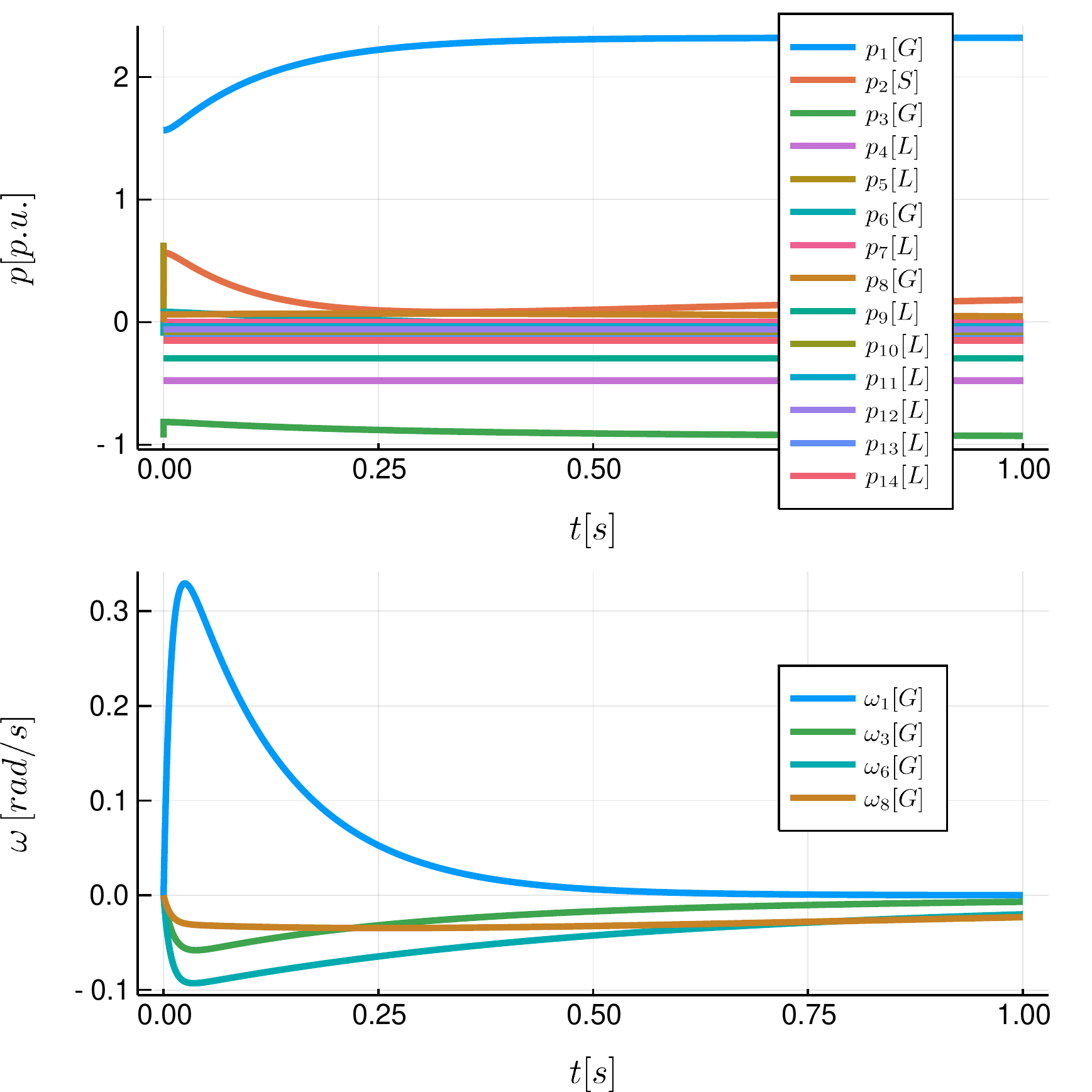}
	\label{fig:ieee14-line-tripping}}
	\hfil
	\caption{The two subfigures show the outflowing power (top) at each node and the angular frequency (bottom) for the buses modeled by the swing equation (1, 3, 6, 8) in for the two perturbation scenarios: (left) frequency perturbation at bus 1 and (right) line tripping of line 2 (between bus 1 and 5). The letter in bracets refers to the modeling type, i.e. $G$ for generator / synchronous compensator as swing equation, $S$ for the slack bus, and $L$ for a load as algebraic PQ-constraint.}
\end{figure*}

Within this section, we will analyze two simple cases: (a) a frequency perturbation at the largest generator at bus 1 and (b) a line tripping event at of line 2 (between bus 1 and 5) (cmp.\ \Cref{fig:ieee14grid}).

Before we find the normal point of operation for the power grid, i.e.\ the fixed point of \Cref{eq:i-definition,eq:u-general,eq:x-general} or synchronous state for the IEEE 14-bus system. For that, we use the grid model \texttt{g} generated in the previous \Cref{sec:implementation} and use the function provided by \pd{}:
\begin{lstlisting}[linewidth=\columnlistingwidth]
fp = operationpoint(g, ones(SystemSize(g)))
\end{lstlisting}
where $\mathtt{ones(SystemSize(g))}$ is a vector of the correct length for the initial condition of the fixed point search. \texttt{fp} is now a \texttt{State} object that we can use as initial condition for the solving the differential equations corresponding to the power grid.

\subsection{Frequency perturbation}

In order to model a frequency perturbation, one can simply take a copy of the fixed point as found before and adjust the initial frequency value:
\begin{lstlisting}[linewidth=\columnlistingwidth]
x0 = copy(fp)
x0[1, :int, 1] += 0.2 
\end{lstlisting}
The second line can be read as ``add 0.2 to the 1$^\text{st}$ internal variable of the 1$^\text{st}$ node'' which is the frequency $\omega$ as there is only one internal variable. Note that the first \texttt{1} refers to the node and then second \texttt{1} to the internal variable counter. The power grid with the initial condition \texttt{x0} can then be solved for a time span of 0.5 seconds by calling:
\begin{lstlisting}[linewidth=\columnlistingwidth]
sol = solve(g, x0, (0.0,.5))
\end{lstlisting}
The solution is shown in \Cref{fig:ieee14-frequency-perturbation}. It shows that the system is very stable against frequency perturbations. The actual dynamics is not so exciting as the system is very stable. Please note that this system was taken as an example to present on how easily one can model a power grid using \pd{}, not to find new exciting dynamics.

\subsection{Line tripping}

To show some more dynamic behavior we simulated a line tripping as well. We model this effect by taking the operation point of the full power grid (\texttt{fp}) as initial condition but defining a new admittance Laplacian where line 2 (between bus 1 and 5, see \Cref{tab:ieee14-line-parameters,fig:ieee14grid}) has been taken out (i.e. the admittance is set to $0$). Running the model with this new Laplacian yields \Cref{fig:ieee14-line-tripping}.

In the frequency plot we identify how the frequency of bus 1, where the line tripping happened, compensates for the momentarily excess power at the bus. The lacking power in the rest of the grid is matched by the synchronous compensators whose frequency decreases in turn. Note that the angular frequency $\omega$ is shown, so with a division by $2\pi$ the maximal frequency deviation $f$ is $\approx 0.05\,$Hz. After about one second, the system recovers to the normal state of operation.

\section{Conclusion \& Outlook}
\label{sec:conclusion}

Within this paper, we have seen how one one can use the Open-Source library \pd{} in order to model the dynamics of a power grid with just a few lines of code. We have seen how the fundamental mathematical equations (given in \Cref{sec:powerdynamics}) translate to source code that reads exactly the same. We employed \pd{} for the IEEE 14-bus distribution grid feeder in order to demonstrate how one can easily simulate faults and analyze the transient reaction of the power grid dynamics. As example scenarios we used a frequency perturbation and a line tripping.

Finally, this paper is really just a sneak preview with a simple example the publication of \pd{} is planned for October 15, 2018. By then, we will have added more inverter control schemes and stochastic descriptions of intermittency due to renewable energy sources.


\section*{Acknowledgment}

This paper was presented at the 19th Wind Integration Workshop and published in the workshop’s proceedings.

We would like to thank the German Academic Exchange Service for the opportunity to participate at the Wind Integration Workshop 2018 in Stockholm via the funding program ``Kongressreisen 2018''.
This paper is based on work developed within the Climate-KIC Pathfinder project ``elena -- electricity network analysis'' funded by the European Institute of Innovation \& Technology.
This work was conducted in the framework of the Complex Energy Networks research group at the Potsdam Institute for Climate Impact Research.

We would like to thank Frank Hellmann and Paul Schultz for the discussions on structuring an Open-Source library for dynamic power grid modeling.



\bibliographystyle{IEEEtran}
%
%
%
\bibliography{windintegration}

\end{document}